# Incidental Interaction: Technology to Support Elder Strength Training through Everyday Movements

Arturo Vazquez Galvez, Christopher Tacca, Isobel Margaret Thompson, Alexander Dawid Bincalar, Christoph Tremmel, Alexander Ng, Richard Gomer, Martin Warner, Chris Freeman, m.c. Schraefel

*Abstract*— Strength training is a key determinant of healthy aging, yet adherence to formal exercise programs among older adults remains low. While many technologies aim to encourage physical activity in older adults, they typically rely on dedicated devices, wearables, or explicit exercise tasks. They therefore do not embed task practice into daily life.

Our new approach, termed Incidental Interaction, instead transforms everyday actions into opportunities for deliberate strength building. It thereby operationalizes everyday movements such as sitting, standing, or lifting objects as strength exercises, encouraging participants to repeat them to build functional capacity. This repetition is encapsulated in the phrase "do it twice", and is combined with movement quality metrics to provide feedback and support progression, without requiring users to adopt new routines or equipment. We illustrate the concept by designing and implementing an ecosystem of instrumented every-day objects and pressure-sensitive mats embedded into ordinary furniture, providing real-time feedback, progress tracking, and motivational cues. To evaluate technical efficacy, we report on two structured pilot deployments with elders (2 week and 4 week studies, n=7).

## I. INTRODUCTION AND RELATED WORK

POPULATIONS worldwide are experiencing rapid demographic shifts: by 2050, one in six people will be over the age of 65, and the number of adults aged 80 years or older is projected to triple to 426 million [3]. Maintaining independence in later life is strongly associated with muscle strength, balance, and mobility [1]. Regular resistance and strength training reduce the risk of falls, hospitalisation, and associated morbidity [4]. Yet, despite these clear benefits, adherence to structured exercise programs among elders is low, with fewer than 30% meeting recommended guidelines. Barriers include lack of time, perceptions of exercise as difficult, and environments that do not naturally encourage practice [2], [5].

Several technologies have attempted to address these challenges, including exergames, physiotherapy robots [6], and wearable activity monitors [7]. While these approaches can be effective in controlled settings, they often require specialised equipment, impose artificial tasks, or fail to integrate naturally into daily life. Reviews of elder-focused fitness interventions consistently report that these limitations undermine long-term adoption. A recent scoping review of 43 interventions found that most systems failed to explicitly address elder-specific design considerations such as compatibility with existing lifestyles, support for dignity and independence, privacy, and emotional engagement [14]. Specifically, the interventions in the scoping review encompassed a range of current fitness technologies for older adults, including wearable activity monitoring devices and smartwatches, screen- or camera-based motion sensing systems used for exercise and rehabilitation, mobile applications prescribing structured exercise routines, and standalone rehabilitation or gym-style equipment designed for home or clinical use. These categories, drawn from the spectrum of technologies evaluated in the selected literature, illustrate that the existing technologies require users to deliberately initiate exercise sessions, interact with dedicated hardware or interfaces, or engage with bespoke software, rather than leveraging actions already embedded in everyday domestic practice. These findings reflect the broader patterns and gaps identified in the literature on how well fitness technologies align with older adults' needs and preferences [14].

Only 16 of the reviewed studies referenced elder-focused design literature, and no intervention scored strongly across all six identified factors: lifestyle fit, similarity to past experience, dignity and independence, privacy, social support, and emotion. Importantly, the review characterises these shortcomings as recurring patterns rather than isolated design oversights. Across intervention types, systems tend to prioritise measurable outcomes or short-term efficacy while introducing novel devices, interaction paradigms, or task structures that remain misaligned with everyday routines and lived experience. As a result, even technically sophisticated interventions often struggle to sustain engagement once deployed in domestic settings. This persistent gap between technological capability and ecological validity remains a central barrier to real-world adoption.

Existing consumer and research systems for promoting physical activity typically rely on coarse goal-setting mechanisms, such as fixed step targets, incremental increases based on prior totals, or population-level comparisons. While effective for general activity tracking, these approaches largely treat movement as an aggregate quantity and rarely incorporate movement quality, temporal structure, or contextual relevance into progression logic. In the context of older adults, such simplified mechanisms risk encouraging inappropriate pacing, overlooking compensatory strategies, or failing to adapt meaningfully to functional ability. As a result, few systems provide principled, adaptive task progression grounded in the structure and quality of everyday movements.

We address these limitations through an "incidental interaction" paradigm that embeds strength training into everyday activities rather than prescribing separate exercise tasks. Specifically, the system integrates sensing into existing household furniture instead of introducing new wearable or dedicated exercise devices, leverages movements that already occur in daily routines (such as sitting, standing, and lifting objects) rather than requiring scheduled exercise sessions, and delivers minimal, context-aware feedback designed to be unobtrusive. Together, these design choices directly support lifestyle compatibility, reduce cognitive and motivational burden, and preserve dignity, independence, and privacy—factors that were consistently under-addressed in prior elder-focused fitness technologies.

This paradigm is based on the observation that many daily movements already constitute strength-demanding activities. Sitting down and standing up or lifting a tin can to a shelf, these are functional strength exercises. Our approach, termed *"do it twice"*, asks elders to repeat such actions when they occur, effectively embedding "reps" into everyday life. It directly addresses several of the criteria that were absent in previous devices: by embedding sensors into ordinary furniture, it achieves lifestyle fit and builds on familiar experiences; by offering unobtrusive feedback, it supports dignity, independence, and privacy; and by framing activity as 'do it twice,' it creates opportunities for enjoyment and motivation. Incidental Interaction advances the state of the art by meeting multiple needs that prior interventions have overlooked.

We illustrate this concept by developing a distributed sensing system integrated into ordinary furniture that detects sit-to-stand actions, classifies repetition structure and movement characteristics, and supports progression through minimal, stage-based feedback. Distinct from existing systems, which typically rely on wearables, single sensing surfaces, or prescribed exercise tasks, this work extends the state of the art in several ways. First, the system captures movement quality and repetition structure using distributed pressure sensing embedded in everyday furniture, without requiring body-worn devices. Second, it supports principled task progression derived from naturally occurring movements rather than predefined exercise routines. Third, the system operationalizes *incidental movement* itself as the primary site of interaction, shifting away from prescribed training toward strength practice embedded in daily life. Forth, by coordinating multiple pressure-sensing surfaces across the environment and an additional peripheral to focus on the upper body, the system enables assessment of combined upper- and lower-body actions that have typically been treated in isolation. Finally, a longitudinal deployment is undertaken to demonstrate how incidental interaction can support strength practice in situ.

II. SYSTEM DESIGN AND IMPLEMENTATION

At a conceptual level, Incidental Interaction is an ecosystem of instrumented objects embedded within the living environment. Everyday artefacts like chairs, armrests, floor mats, and handheld items, are augmented with sensing surfaces that unobtrusively capture strength-related movements without requiring new or artificial tasks. Together, these objects form a distributed infrastructure that transforms routine actions into structured strength practice. The concept of *incidental interaction* originates from Dix [13], who describes interactions that occur as a by-product of ongoing activity rather than through deliberate, goal-directed engagement. In this framing, "incidental" refers to effects that emerge without requiring users to explicitly decide to interact with a system. Prior work has used this concept to describe how digital systems can leverage routine behaviour, attention, and context to shape interaction without imposing additional cognitive or motivational burden. In this work, we extend incidental interaction to the domain of physical activity by treating everyday bodily movements themselves as interaction opportunities. Rather than asking users to initiate exercise or engage with a dedicated interface, the system responds to actions that are already occurring as part of daily life. This perspective aligns with psychological accounts of habit formation and situated action, which emphasise that behaviours embedded in routine contexts are more likely to be sustained than those requiring explicit intention or self-regulation.

An example is shown in Figure *1*, where the ecosystem provides complementary perspectives on movement (sit-to-stand transitions, balance support, upper-limb lifting) while delivering coherent multimodal feedback through a touchscreen interface, designed to motivate, guide technique, and summarise progress.

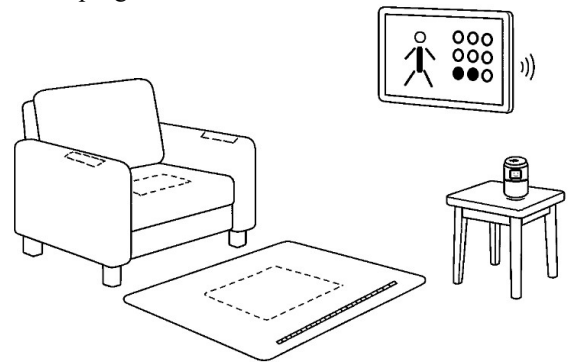

Figure 1. Conceptual diagram of a room with an ecosystem of instrumented objects for incidental interaction.

Creating such an integrated platform raises significant technical challenges: developing textile-based pressure grids that are both durable and sensitive, calibrating across diverse body types and movement styles, ensuring reliable wireless communication among distributed microcontrollers, and simplifying the feedback interface for users with limited digital literacy. By explicitly addressing these challenges, our approach demonstrates that it is possible to transform familiar furniture and objects into a distributed infrastructure for strength training, aligning ecological validity with ambitious technical innovation.

*2.2 Wireless Communication Topology*

All embedded sensors communicate with a central hub using Bluetooth Low Energy (BLE). The topology follows an IoT-style star network, where lightweight peripherals (seat cushion, armrests, floor mat, handheld objects) transmit data packets to a coordinating hub for aggregation and analysis. Each peripheral is built around a low-power wireless microcontroller, following a common modular architecture that integrates sensing, local pre-processing, and wireless communication. Figure 2 illustrates the topology, where multiple peripherals operate concurrently.

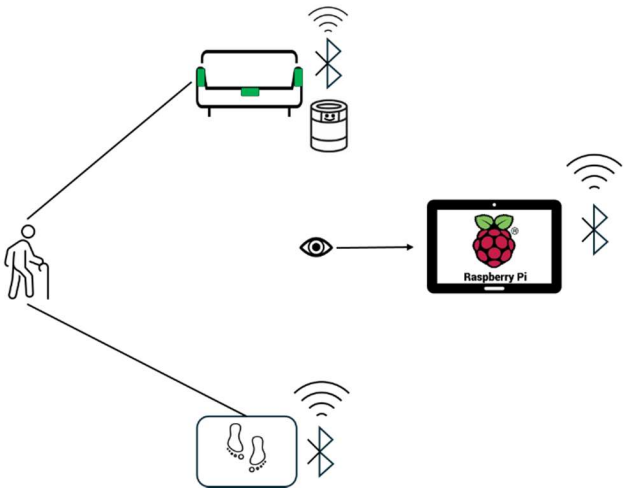

*Figure 2. Wireless communication topology linking peripheral sensors to the central hub*

The hub buffers incoming streams, timestamps packets, and immediately drives the touchscreen output. This design supports a larger number of simultaneous sensing surfaces, tolerates irregular event frequencies, and minimises packet loss. These features are crucial in domestic environments, where wireless interference and unpredictable usage patterns are the norm.

At a system level, the hub maintains a dynamic registry of connected sensing modules, allowing individual devices to join or leave the network without requiring manual reconfiguration. Each module periodically transmits identifying metadata alongside sensor data, enabling the hub to associate incoming packets with specific objects or sensing surfaces and manage multiple concurrent streams. Incoming data are buffered and timestamped on receipt, which allows the system to tolerate asynchronous transmissions, variable event frequencies, and temporary communication delays across modules. Together, these mechanisms support scalable deployment of multiple peripherals while maintaining robust operation in domestic wireless environments.

### 2.3 Sensors & Hardware

All sensing modules share a common hardware architecture built around an ESP32 microcontroller, which integrates the BLE communication, on-board processing, and flexible I/O for sensor interfacing. Power is supplied via a rechargeable lithium-ion battery with on-board battery management circuitry. This shared architecture enables consistent firmware, communication protocols, and power handling across modules, while allowing individual sensing surfaces to differ in form factor, sensor layout, and sampling characteristics.

- The floor mat module consists of a pressure-sensitive textile matrix embedded within a flexible mat sized to support standing balance and weight-shifting activities. The sensing surface captures distributed pressure patterns during stance and transitions, allowing detection of foot placement. The centre of pressure (CoP) of each foot is computed using highly efficient algorithms detailed in [15]. These values are then sampled at a fixed rate and transmitted wirelessly to the central hub, where temporal alignment with other modules is performed.
- The armrest (arm mat) module integrates pressure sensing into padded arm supports, capturing upper-limb loading and support behaviour during sit-to-stand transitions and seated activity. This placement enables detection of asymmetric loading and compensatory strategies without requiring body-worn sensors. The module uses the same embedded controller and communication pipeline as other sensing surfaces, supporting consistent integration within the distributed system.

The seat cushion module functions as the primary reference sensor for sit-to-stand activity within the system. A pressure-sensitive textile grid embedded in the cushion detects sustained load corresponding to seated posture, allowing the system to reliably distinguish between sitting and standing states without requiring body-worn sensors. Transitions between these states form the basis for identifying sit-to-stand events, which are used to structure repetition counting (singles versus doubles), timing, and progression logic. By treating absence of pressure as an explicit standing state rather than a sensor failure, the cushion provides a simple and robust mechanism for anchoring activity detection across the wider sensing ecosystem.

- The handheld module ("Can Band") is designed to capture grasping and lifting actions. Figure 4 shows an initial prototype, while the final version will consist of a band that is placed around an existing can or object. It integrates a compact pressure-sensitive surface (to sense the grip force), an inertial measurement unit (to detect the direction, duration and length of each movement), a light level sensor (to trigger sleep mode when placed in a cupboard), and an LED array (to provide visual instructions and feedback). This module prompts the user to repeatedly perform upper-limb activities, extending the sensing ecosystem beyond static furniture surfaces. As with other modules, local sampling and pre-processing are handled on-device before wireless transmission to the hub.

Figure 3 shows the pressure-sensitive grids fabricated with conductive electrodes and a polymeric foil which are embedded into the cushions, armrests, and floor mats. As indicated LEDs are integrated into the floor mats to provide immediate visual confirmation of activation. This feedback aids calibration, reassures users, and assists researchers in verifying sensor operation during deployment. However, embedding sensors in soft materials introduces durability challenges. To mitigate this, the grids are reinforced with protective textiles, and calibration routines are included in software.

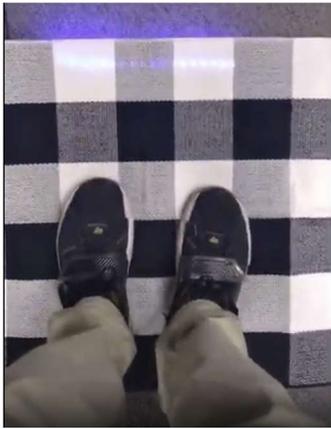

Figure 3. LED lighting visible along top edge, indicating activation of floor mat sensor.

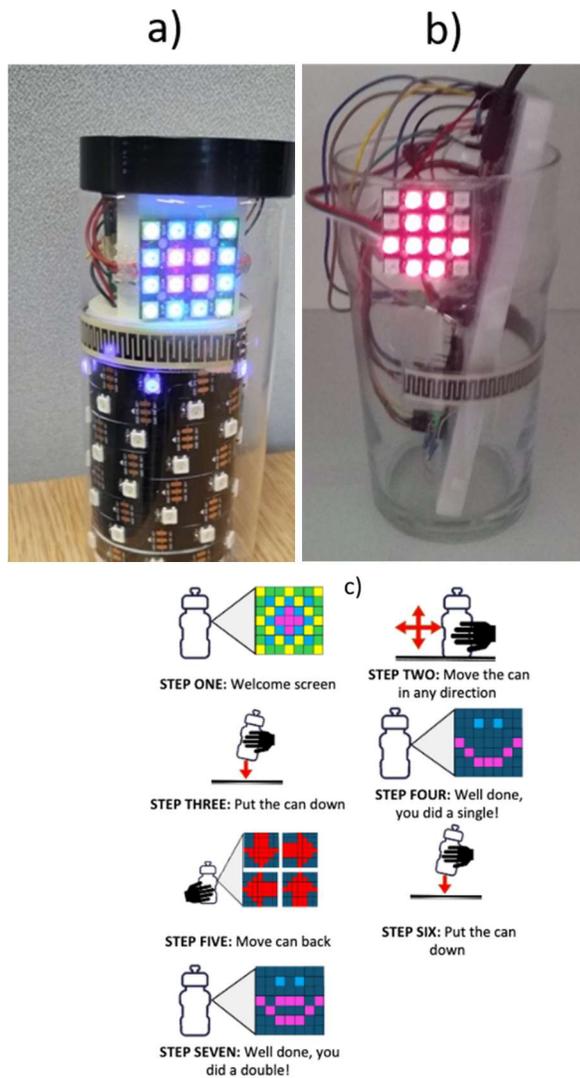

*Figure 4. Can Band for upper-limb interaction device.*
*(a) Hardware prototype showing embedded electronics and LED matrix. (b) Can Band attached to a drinking glass, illustrating an example illumination pattern from the embedded LED array used to cue put down object after lifting. (c) Schematic of the illumination phases and corresponding lift–hold–return movement sequence. Different LED patterns are used to guide users through each phase of the action.*

## 2.4 Signal Processing Pipeline

The signal processing pipeline is designed to operate across all sensing modules in the ecosystem, including static pressure surfaces (seat cushion, armrests, floor mat) as well as handheld sensing devices such as the Can Band. The hub executes a multi-stage signal-processing pipeline that converts raw sensor signals into activity classifications and metrics.

Figure 5 shows the pipeline for sit-to-stand cycles. The detection stage distinguishes between single and double repetition by applying three criteria:

1. **Temporal thresholds:** Each sit-to-stand must occur within a minimum and maximum duration set between 15–20 sec depending on calibration. Movements exceeding this range are excluded to filter out incidental weight shifts or prolonged partial stands.
2. **Pressure distribution:** Weight transfer must shift from seat to mat above defined percentage thresholds and maintain it for a very small-time (between $0.5 – 2$ seconds) window that ensures actual rising rather than repositioning.
3. **Sequence integrity:** If a pause longer than the designated time for a double occurs between rising and sitting phases, the sequence resets, and no repetition is logged.

Each valid repetition is timestamped and stored with the following metadata: repetition type (single or double), execution duration, sensor signature, and time of day. Invalid sequences are discarded.

Following the detection stage, the metric stage calculates duration, symmetry, and cycle-to-cycle consistency. A key design challenge was balancing accuracy with computational efficiency: the hub must process streams from multiple peripherals simultaneously while avoiding perceptible delay in feedback. This architecture ensures that movements are not only counted but also characterised with sufficient fidelity to support training insights. While individual modules differ in sensing geometry and placement, the same pipeline, comprising signal smoothing, temporal segmentation, event detection, and repetition classification is applied to both environmental pressure surfaces and handheld devices, enabling consistent interpretation of sit-to-stand and upper-limb actions within a unified framework.

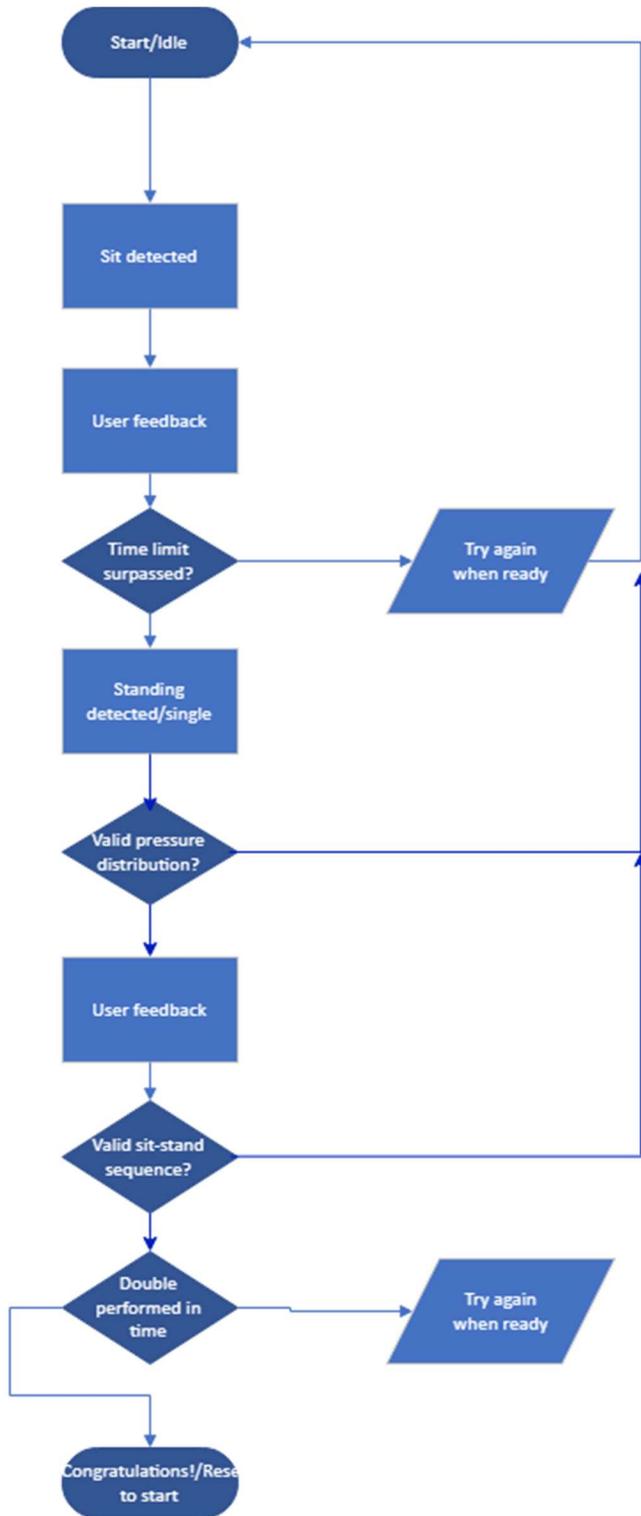

*Figure 5. Sit-to-stand repetition detection and validation flow. The pipeline incorporates temporal thresholding, pressure distribution validation, and sequence integrity checks to distinguish valid single and double repetitions before computing performance metrics and triggering user feedback.*

*2.5 Task Progression Algorithm*

Incidental interaction devices operate by delivering cues to encourage specific physical actions; providing real-time feedback during movements to guide execution; and presenting summary feedback on movement quantity and quality over daily and weekly timescales. Precise selection and timing of these cues and feedback are critical to maximise both short- and long-term effectiveness while avoiding fatigue, frustration, or overexertion. By coupling progression to validated movement patterns and observed engagement, the system allows goals to adapt gradually and, over time, to recede as practices become embedded into everyday routines. This contrasts with existing elder-focused technologies, such as FitBit or Apple Watch, whose progression mechanisms remain largely coarse-grained, relying on simple event counting, static targets, simplistic comparisons, population-level comparisons, or uniform increases that are insensitive to functional movement quality.

The task progression algorithm ties progression to the structure and quality of individual movements as they occur in daily life, and is grounded in a points-based model that extends the goal-oriented approaches used in commercial activity trackers. The algorithm integrates temporal constraints, pressure distribution features, and adaptive progression logic to ensure that only valid, functionally meaningful movements are recognised and that goals evolve according to actual user engagement.

To make this progression logic explicit, the system implements a simple weekly goal-evaluation rule based on recent user performance. Rather than relying on continuous optimisation or complex weighting schemes, progression is driven by consistency in achieving the current daily target.

Weekly goal update rule. Let $G$ denote the current daily double target. For each day $i \in \{1, \ldots, 7\}$, let $D_i$ be the number of sit-to-stand double repetitions performed on day $i$. We compute:

$$N = \sum_{i=1}^{7} 1(D_i \geq G)$$

Where $1(\cdot)$ is an indicator function. If $N \geq 3$, the system prompts the user to increase their daily goal for the following week; otherwise, the goal remains unchanged. In addition, the mean execution time of double repetitions is displayed to users as performance feedback but is not used to trigger goal increases in the current implementation.

*2.6 Daily Metrics and Weekly Goal Evaluation*

For each day, the algorithm maintains three core metrics, and all data are stored locally:

- **Singles count (S)**: total valid single repetitions.
- **Doubles count (D):** total valid double repetitions.
- **Double execution times (T):** list of execution durations and timestamps for doubles, used to compute daily average performance time and temporal activity distribution (e.g., morning vs evening usage).
- **Balance leaning:** If one or both of the arm rest sensor was used when performing a double.
  **Time of day:** the specific time of day when activity is detected.

In addition, the Can Band module contributes the following upper-limb metrics:

- **Single lifts count ($S_{cb}$)**: total number of valid single lifting actions detected per day.

- **Double lifts count ($D_{cb}$)**: total number of valid double lift–return sequences detected per day.
- **Double execution time $T_{cb}$** : time taken to complete a full lift–hold–return cycle with the object.
- **Movement distance ($D_{cb}$)**: vertical/horizontal displacement of the object during a double, estimated from inertial measurements.
- **Grip strength ($F_{cb}$)**: average and peak grip force applied during lifting, measured in Newtons using the pressure-sensitive surface.

These metrics are logged alongside sit-to-stand measures and displayed to users as performance feedback, but are not currently used to drive automatic goal progression.

Every seven days, the algorithm runs a progression check routine to evaluate whether the user's current daily goal remains appropriate:
- G = current daily double goal
- D = number of doubles achieved on day i (i = 1…7)
- N = number of days during the week where D > G

The algorithm counts how many days the user has exceeded their daily goal. If N ≥ 3, the system generates a prompt on the touchscreen encouraging the user to increase their daily target. For example, it might display: "Great job! You've exceeded your daily goal 3 times this week. Would you like to set a higher goal for next week?" The user can either accept the increase (incrementing G by a fixed step or user-specified value) or keep their current goal. If N < 3, the system displays a motivational message reinforcing continued engagement without suggesting a goal change. Figure 6 shows the touchscreen interface used to present daily goal progress and access weekly summaries for sit-to-stand activity, while Figure 7 illustrates the LED-based goal and progress feedback provided directly on the Can Band device.

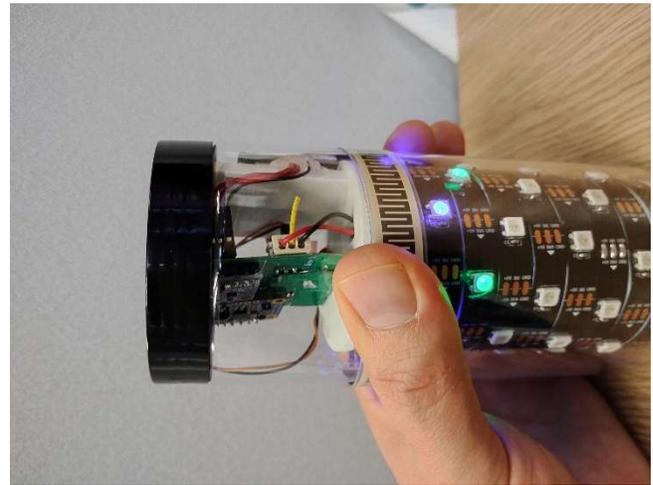

*Figure 7 multi-colour LED array mounted around the object provides real-time cues for movement phase (lift, hold, return) and repetition completion, serving as the primary goal and progress feedback mechanism for upper-limb exercises.*

### 2.7 Feedback Modules

Outputs from the processing pipeline are reflected in several forms of guidance. A screen displays the current movement stage, update repetition counters, and animate progress bars. Weekly summaries present cumulative repetitions, average cycle durations, and adherence trends. Additional modes include pause control, which suspends logging during social contexts, and goal-masking, which hides counts to discourage fixation on short-term numbers.

Figure 8 shows all stages consolidated into a single view. Each screen corresponds to a distinct module of the system: stage guidance is tied to the task progression algorithm; repetition counters are updated by the detection module; weekly summaries are derived from aggregated metrics.

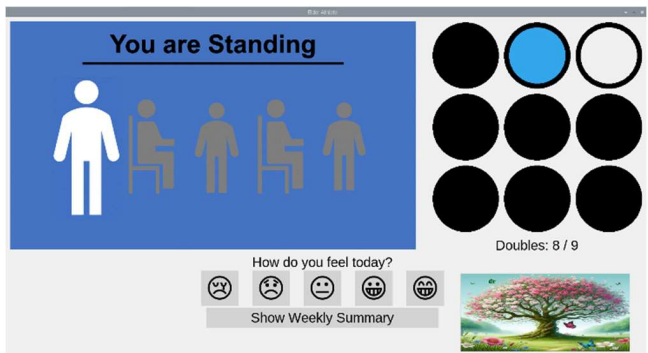

*Figure 6 Goal progress and feedback interface for sit-to-stand activity The touchscreen display shows real-time movement state, daily double progress toward the current goal (circles), and access to weekly summaries and subjective feedback*

For the Can Band device, goal progress and task guidance are communicated directly on the device using an embedded multi-colour LED array. These values are also listed in tabular form on the touchscreen interface. The LEDs indicate double repetitions in real time, providing immediate feedback during object interaction.

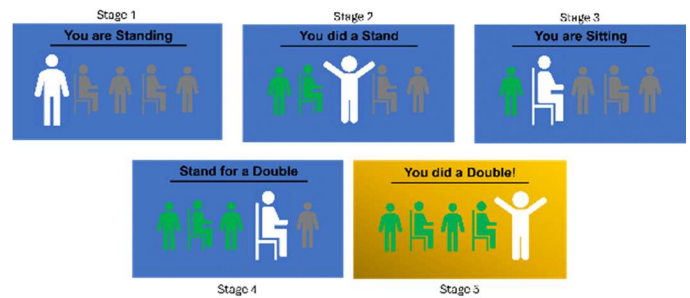

*Figure 8. Sitting-to-standing stages. Five stages of motion detection displayed to the user in real time.*

By presenting feedback in this structured way, the system links sensing, processing, and guidance into a coherent loop. This modularity allows new feedback strategies to be added without altering the sensing hardware, supporting extensibility in future deployments.

### III. METHODS FOR EVALUATION AND PRELIMINARY FINDINGS

The system was deployed with older adult participants to validate the technical efficacy of the sensing, communication and feedback modules in real-world conditions.

The study was approved by the University of Southampton ethics committee. Two pilot deployments were conducted: one lasting two weeks and one lasting four weeks. A total of seven older adults were recruited through care homes and community partners. All participants provided informed consent prior to participation, and they ranged in age from 56 to 90 years and included both male and female participants. Participants were asked to engage with the system as part of their everyday routines within their home environments.

During both deployments, all peripherals (seat cushion, armrests, floor mat, handheld Can Band) remained stable across daily use, with negligible packet loss observed in logged transmissions. The signal-processing pipeline successfully segmented sit-to-stand cycles, and the task progression algorithm correctly classified single and double repetitions without manual recalibration once initial setup was complete. Feedback interfaces updated in real time, and participants consistently reported that system responses aligned with their actions. In summary, the pilots confirmed:

- **Hardware robustness** under repeated use (textile sensors maintained function).
- **Stable communication** across distributed peripherals.
- **Accurate classification** of sit-to-stand events.
- **Responsive feedback** aligned with user actions.

These confirm that the system can operate continuously for weeks, maintain stable BLE transmission under irregular usage, and deliver real-time feedback. The quality-control logic successfully filtered incomplete or unstable repetitions, ensuring that only valid actions were logged. The findings validate the technical capability of the system in situ, providing the foundation for future studies that may target clinical or behavioural outcomes.

Figure 9, shows the full system setup, including the floor, cushion and arm rests sensors. Facing these is the touch screen that displays feedback to the user.

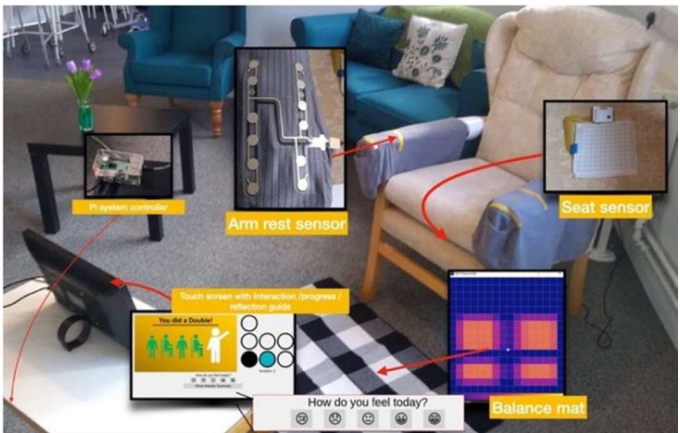

*Figure 9. Prototype in use. Example of system setup in front of a chair with floor mat and cushion sensor.*

## IV. Discussion

This discussion focuses on the technical performance and deployment characteristics of the system rather than on evaluating the effectiveness of the interaction paradigm itself. The current study was scoped to validate sensing accuracy, signal processing, wireless communication, and real-time feedback under realistic home conditions.

From a technical perspective, three insights stand out. First, sensor calibration was required to adapt to diverse body sizes, seating preferences, and movement styles. Second, the feedback interface requires further simplification for users with limited digital literacy, with large fonts, clear icons, and minimal menus proving essential. Third, reliable segmentation of sit-to-stand cycles depends on careful tuning of temporal thresholds and pressure-distribution parameters across deployments.

Technical failures (resets, wires, sensor malfunctions) discouraged use. The fixed installation limited flexibility, and intrusiveness in social contexts reduced usability in shared domestic environments.

## V. Future Work

Building on these feasibility insights, future work will prioritize technical refinements alongside broader deployment studies. Hardware simplification is critical: reducing wiring, developing fully wireless sensor modules, and ensuring easy portability between chairs or environments. Reliability improvements include automatic recovery from disconnection, self-calibration routines, and edge processing for reduced latency. Motivational features will extend beyond novelty, incorporating gamification, progress tracking, and potentially light social competition or group challenges [12].

Building on these feasibility insights, future work should prioritize:

1. **Longer and larger trials** to evaluate sustained impact on strength, mobility, and fall risk.
2. **Simplifying hardware**: wireless sensors, fewer wires, portable or movable designs.
3. **System reliability**: enabling independent resets, ensuring automatic sleep/power management.
4. **Enhancing engagement**: gamified feedback, progress badges, social support features, and personalization to maintain motivation beyond novelty.
5. **Clinical integration**: aligning with falls-prevention programs and healthy ageing initiatives, including potential links to rehabilitation and primary care.

## VI. Conclusions

This paper presented a distributed sensing system that embeds pressure and inertial sensing into everyday furniture and objects to detect and characterise strength-related movements in the home. Through two in-home deployments, we demonstrated that the system can reliably identify sit-to-stand and object-lifting actions, distinguish singles and doubles, compute movement metrics, and deliver real-time feedback over multi-week periods.

The results show that such sensing and processing pipelines can operate under realistic domestic conditions, supporting continuous data capture and interaction without requiring body-worn devices or dedicated exercise equipment. These findings establish the technical feasibility of using embedded, environment-level sensing to support the study and design of strength-training interactions in everyday settings.